\documentstyle[12pt]{article}
\newcommand{\proof}[1]{{\bf Proof:} #1~$\Box$.}
\newcommand{\bb}{\begin{equation}}
\newcommand{\ee}{\end{equation}}
\newcommand{\bqn}{\begin{eqnarray}}
\newcommand{\eqn}{\end{eqnarray}}

\newtheorem{theorem}{Theorem}

\begin{document}
\begin{titlepage}

\begin{flushright}

ULB--TH--97/02\\
hep-th/9702042\\

\end{flushright}
\vfill

\begin{center}
{\Large{\bf  BRST cohomology of the Chapline-Manton
model}}
\end{center}
\vfill

\begin{center}
{\large
Marc Henneaux$^{a,b}$,
Bernard Knaepen$^{a,*}$ \\ and
Christiane Schomblond$^a$}
\end{center}
\vfill

\begin{center}{\sl
$^a$ Facult\'e des Sciences, Universit\'e Libre de
Bruxelles,\\
Campus Plaine C.P. 231, B--1050 Bruxelles, Belgium\\[1.5ex]

$^b$ Centro de Estudios Cient\'\i ficos de Santiago,\\
Casilla 16443, Santiago 9, Chile

}\end{center}
\vfill

\begin{abstract}
We completely compute the local BRST cohomology 
$H(s\vert d)$ of the
combined Yang-Mills-2-form system coupled through the
Yang-Mills Chern-Simons term (``Chapline-Manton model").  We
consider
the case of a simple gauge group and 
explicitely include in the analysis the sources for the
BRST variations of the fields (``antifields").
We  show that there is an  antifield independent
representative in each cohomological class of 
$H(s\vert d)$ at ghost
number
$0$ or $1$.
Accordingly, any counterterm may be
assumed to preserve the gauge symmetries.  Similarly,
there is no new candidate anomaly beside those already
considered in the literature, even
when one takes the antifields into account.
We then 
characterize explicitly all the non-trivial
solutions of the Wess-Zumino consistency
conditions.  In particular, we provide
a cohomological interpretation of the Green-Schwarz
anomaly cancellation mechanism.

\end{abstract}

\vspace{5em}

\hrule width 5.cm
\vspace*{.5em}

{\small \noindent (*)Aspirant du Fonds National de la
Recherche Scientifique (Belgium).}

\end{titlepage}

\section{Introduction}
\setcounter{equation}{0}
\setcounter{theorem}{0}

Chern-Simons couplings  of two-forms fields to  Yang-Mills 
gauge fields
play a central role in the Green-Schwarz anomaly
cancellation mechanism \cite{GS} and, for this reason, are
important in string theory \cite{GSW}. In this letter we
completely work out the general solution of the Wess-Zumino
consistency
condition at all ghost numbers (BRST cohomology 
$H(s\vert d)$) in
the space of local exterior $n$-forms depending on the 
fields and the
antifields, for the Chapline-Manton
model whose Lagrangian reads
\cite{ChaplineManton,Nito,Cham,BergRooWitNieu},
\begin{eqnarray}
{\cal L}&=&-\frac{1}{4}F_{\mu\nu}^a F_a^{\mu\nu}
-\frac{1}{12} H_{\mu\nu\rho}H^{\mu\nu\rho},
\label{Lagrangian} \\ F^a_{\mu\nu}&=&\partial_\mu A^a_\nu -
\partial_\nu A^a_\mu - C^a_{bc}A^b_\mu A^c_\nu, \\
H_{\mu\nu\rho}&=&\partial_\mu B_{\nu\rho} + \partial_\nu
B_{\rho\mu} + \partial_\rho B_{\mu\nu} - 2\lambda
C_{abc}A^a_\mu A^b_\nu A^c_\rho \nonumber \\&& -
2\lambda(A^a_\mu F_{a\nu\rho} +A^a_\nu F_{a\rho\mu}+A^a_\rho
F_{a\mu\nu}) \label{Hmunurho}.
\end{eqnarray}
Here $A^a_\mu$ is the Yang-Mills vector potential,
$B_{\mu\nu}$ is an abelian two-form and $\lambda$ is the
coupling constant. The $H_{\mu\nu\rho}$ are the
components of the exterior form $H=dB+ \lambda \omega_3$ 
where
$\omega_3=tr(AF+\frac{1}{3}A^3)$ is the Chern-Simons
three-form. For definiteness, the gauge group G is
taken to be simple. 

Our main result is that the antifields can be
completely eliminated in cohomology at ghost number zero
and one. This implies that:
\begin{itemize}
\item
there are no new
antifield-dependent anomalies and the only possible
anomalies of the coupled system are those of the pure
Yang-Mills theory that are not made trivial by the
coupling to the two-form;
\item
the
counterterms may always be chosen so as to preserve
the gauge symmetries.  Thus, structural constraints of the
type considered in \cite{GomisWeinberg} can be consistently
imposed.
\end{itemize}

In order to establish this result, we follow the approach 
developed in
\cite{BBH2} for the pure Yang-Mills theory.  This is made 
possible by a
change of variables that brings the BRST differential $s$ 
to the same form
$s = \delta + \gamma$ (without extra higher order 
contributions).  Here,
$\delta$ is the Koszul-Tate differential associated with 
the gauge
covariant equations of motion while $\gamma$ is, up to 
inessential terms, 
the coboundary operator of the Lie algebra cohomology
of $G$ in some definite representation space (sections 2 
and 3).

Once $s$ has been brought to the form $s = \delta + 
\gamma$, the
computation of the cohomology $H(s\vert d)$ of $s$ modulo 
the
spacetime exterior derivative $d$ proceeds by expanding 
the cocycles
according to the antighost number.  The obstructions for 
removing the
antifields from a given cocycle lie in the groups 
$H(\delta \vert d)$ of
the
``characteristic cohomology",  as in the pure Yang-Mills 
case.
We thus compute $H(\delta \vert d)$, which has been 
related in
\cite{BBH1} to the conservation laws of first and higher 
orders.
We find that the conservation laws for the Chapline-Manton 
model are of
only two
types: conserved tensor $H^{\lambda \mu \nu}$ of rank three,
$\partial_\lambda H^{\lambda \mu \nu} \approx 0$, and 
ordinary
conservation laws of rank one associated with rigid 
symmetries
(e.g., Poincar\'e symmetries).  The corresponding 
obstructions cannot
arise at ghost number zero or one, for which one can 
accordingly
always remove the antifields by adding exact terms 
(section 4).

The other important ingredient of the analysis is that all 
the solutions
of the Wess-Zumino consistency conditions involving $tr C^3
\equiv
C_{abc} C^a C^b C^c$ (where $C^a$ is the Yang-Mills ghost) 
and
$tr F^2$, or related
to them through the descent
equations, are removed from the BRST cohomology.   This is 
because $trC^3$ is
$s$-exact when the coupling is turned on, $trC^3 = 
(3/\lambda) s\rho$, where
$\rho$ is the ghost of ghost associated to the reducibility
of the 2-form gauge symmetries.  In the same way, $trF^2$ 
is $d$-exact in
the space
of gauge invariant exterior forms, $trF^2 = (1/\lambda) 
dH$.  These
properties are
at the core of the Green-Schwarz anomaly cancellation 
mechanism 
and are discussed in section 5.

\section{BRST differential}
\setcounter{equation}{0}
\setcounter{theorem}{0}

According to the standard rules of the BRST formalism
\cite{BRST,BV,HenneauxTeitelboim}  we introduce, beside the
fields ($A^a_\mu, B_{\mu\nu}$), the antifields $(A^{*\mu}_a,
C^*_a,B^{*\mu\nu},
\eta^{*\mu},
\rho^*)$ and the ghosts $(C^a, \eta_\mu, \rho)$, with
Grassmann parity given by,
\begin{equation}
\epsilon(A^{*\mu}_a)=\epsilon(B^{*\mu\nu})=
\epsilon(\rho^*)=1;
\ \epsilon(C^{*}_a)=\epsilon(\eta^{*\mu})= 0.
\end{equation}
The action of the
BRST differential
$s$ on the algebra ${\cal P}$ of spacetime forms with
coefficients that are polynomials in the fields,
antifields, ghosts and their derivatives is defined
through \cite{BV,HenneauxTeitelboim}
\begin{eqnarray}
sA^a_\mu &=&\partial_\mu C^a - C^a_{bc} A^b_\mu C^c=D_\mu
C^a,
\\ sC^a&=&\frac{1}{2} C^a_{bc} C^b C^c, \\
sB_{\mu\nu}&=&2\lambda C_a(\partial_\mu A^a_\nu
-\partial_\nu A^a_\mu) +\partial_\mu \eta_\nu - \partial_\nu
\eta_\mu, \\ s\eta_\mu &=& \lambda C_{abc}C^a C^b A^c_\mu
-\partial_\mu \rho,\\ s\rho &=& \frac{1}{3}\lambda
C_{abc}C^a C^b C^c, \\
sA^{*\mu}_a &=& D_\nu F_a^{\nu\mu}+2\lambda
H^{\mu\nu\rho}F_{a\nu\rho} -
2\lambda\partial_\rho H^{\rho\mu\nu}A_{a\nu}\nonumber \\
&&- 2\lambda\partial_\mu (B^{*\nu\mu}C_a)
-\lambda\eta^{*\mu}C_{abc} C^b C^c + C_{abc} A^{*b\mu}C^c,
\label{sastar} \\ sC_a^* &=&2\lambda
B^{*\mu\nu}\partial_\mu A_{a\nu}+2\lambda
C_{abc}\eta^{*\mu}C^b A^c_\mu +
\lambda C_{abc}\rho^* C^b C^c \nonumber \\ && - D_\mu
A^{*\mu}_a - C_{abc}C^{*b} C^c,
\\
sB^{*\mu\nu}&=&\partial_\rho H^{\rho\mu\nu}, \\
s\eta^{*\mu}&=&-\partial_\nu B^{\nu\mu}, \\
s\rho^* &=& -\partial_\mu \eta^{*\mu}.
\end{eqnarray}
The action of $s$ on the fields and the ghosts takes a 
clearer
form when rewritten in differential form notations,
\begin{eqnarray}
sA+DC=0, \\
sC=C^2, \\
sB+ \lambda \omega_2 + d\eta =0, \\
s\eta + \lambda \omega_1 + d\rho =0, \\
s\rho = \frac{1}{3}\lambda tr C^3.
\end{eqnarray}
Here, the one-form $\omega_1$ and the two-form $\omega_2$ 
are
related to the Chern-Simons form $\omega_3$  through the 
descent,
\begin{eqnarray}
s\omega_3 + d \omega_2&=&0, \ \ \omega_2= tr (CdA), \\
s \omega_2 +d\omega_1&=&0, \ \ \omega_1= tr (C^2 A),\\
s \omega_1 + d(\frac{1}{3} tr C^3)&=&0.
\end{eqnarray}

In terms of the above variables,
the BRST differential has two major defects. The
first is that it has a component of antighost number 1.
There are indeed terms of ``higher
order" in $s$ \cite{HenneauxTeitelboim},
e.g. $\eta^{*\mu}C_{abc} C^a C^b$ in
(\ref{sastar}).  Consequently, the BRST
differential
does not split
as the sum of the Kozsul-Tate differential and the
longitudinal exterior derivative as it does when
the fields are not coupled. 
The second undesired feature is that the
BRST variations of the antifields of the Yang-Mills sector
contain contributions not covariant  under the gauge
transformations, e.g. $\partial_\rho H^{\rho\mu\nu}
A_{a\nu}$ in (\ref{sastar}). One can remedy
both problems by redefining the
antifields of the Yang-Mills sector according to the
following invertible transformations:
\begin{eqnarray}
A^{*\mu}_a \rightarrow A^{*\mu}_a + 2 B^{*\mu\nu}A_{a\nu} -
2 \eta^{*\mu}C_a,\label{var1} \\ \label{var2}
C^*_a \rightarrow C^*_a + 2 \eta^{*\mu}A_{a\mu}-2\rho^*C_a.
\end{eqnarray}
In terms of the new variables, the BRST differential
takes the familiar form,
\begin{equation}
s=\delta + \gamma,
\end{equation}
with:
\begin{eqnarray}
\delta B^{*\mu\nu}=\partial_\rho H^{\rho\mu\nu};\ \
\delta \eta^{*\mu}=-\partial_\nu B^{*\nu\mu};\ \
\delta\rho^* = -\partial_\mu \eta^{*\mu}; \label{delta1} \\
\delta A^{*\mu}_a = D_\nu F_a^{\nu\mu}+2\lambda
H^{\mu\nu\rho}F_{a\nu\rho};\ \
\delta C_a^* =2\lambda B^{*\mu\nu}F_{a\mu\nu} - D_\mu
A^{*\mu}_a, \label{defdelta}
\end{eqnarray}
and
\begin{equation}
\gamma B^{*\mu\nu}=\gamma \eta^{*\mu}=\gamma \rho^* = 0;\ \
\gamma A^{*\mu}_a = C_{abc}A^{*b\mu}C^c;\ \ \gamma C_a^* =
-C_{abc}C^{*b}C^c \label{delta2};
\end{equation}
\begin{equation}
\gamma \ (fields) = s \ (fields).
\end{equation}
The
$\gamma$ variations of the Yang-Mills variables are
now identical to those of the uncoupled theory and show
that $A^{*a}_\mu$ and $C^*_a$ transform according to the 
adjoint
representation.

\section{Cohomology of $\delta$ and $\gamma$}
\setcounter{equation}{0}
\setcounter{theorem}{0}

\subsection{$H(\delta)$}
The new antifields defined by (\ref{var1}) and
(\ref{var2}) are no longer homogenous in the standard 
antighost number,
since they mix antifields with different antighost numbers.
We thus redefine the antighost number as
\begin{eqnarray}
antigh (fields) = antigh (ghosts) = 0; \label{anti1}\\
antigh (A^{*\mu}_a)= 1;\ \ antigh (C^{*}_a)= 2;  
\label{anti2}\\
antigh (B^{*\mu\nu})= 1;\ \ antigh (\eta^{*\mu})= 2;  \
\ antigh (\rho^*)=3. \label{anti3}
\end{eqnarray} 
In terms of the new antighost
number, the differential
$\delta$ defined by  (\ref{delta1}) - (\ref{delta2}) has 
antighost -1
and the differential $\gamma$ has antighost number 
zero. From
now on, the antighost number
will always refer to (\ref{anti1}) -
(\ref{anti3}). 

The initial equations of motion
are not manifestly covariant under the internal
gauge symmetries since they contain a
``bare" $A$.  The redefinition (\ref{var1}) - (\ref{var2}) 
replaces these
equations by linear combinations of them which are, by
contrast,
manifestly covariant.  The new equations are clearly
equivalent to the original ones and are obtained by 
equating to zero the
$\delta$-variations of the new antifields
of antighost number 1,
\begin{equation}
\partial_\rho H^{\rho\mu\nu} = 0, \; \;
D_\nu F_a^{\nu\mu}+2\lambda
H^{\mu\nu\rho}F_{a\nu\rho} = 0.
\end{equation}
Similarly, the reducibility identities on the equations of 
motion
encoded in the Koszul-Tate variations of the antifields of
antighost number 2 are also manifestly invariant under gauge
transformations.  For this reason, one can call $\delta$ the
``covariant Koszul-Tate differential".  Because $\delta$ 
encodes
a complete set of equations of motion and  reducibility
identitites, one has the standard result,
\begin{theorem}
$H_i(\delta)=0$ for $i>0$, where $i$ is the antighost
number, i.e, the cohomology of $\delta$ vanishes in
antighost number strictly greater than zero.
\label{propkoszul}
\end{theorem} In degree zero, the
cohomology of $\delta$ is the algebra of ``on-shell
functions"
\cite{FHST,FH,HenneauxTeitelboim,HenneauxCMP}.

\subsection{$H(\gamma)$}
One also associates to $\gamma$ another grading called
the `pureghost number', which is given by,
\begin{eqnarray}
puregh (fields) = puregh (antifields) = 0; \\
puregh (C^a)= 1;\ \ puregh
(\eta_{\mu})= 1; \
\ puregh (\rho)=2.
\end{eqnarray}
The `ghost' number is then defined as the
difference between the pureghost number and the antighost
number,
$gh=puregh -antigh$.

When $\lambda=0$ (uncoupled case), the cohomology of
$\gamma$, $H(\gamma)$, is given by the tensor product of
the pure Yang-Mills cohomology and of the free 2-form
cohomology which have already been calculated separately
in
\cite{CohoH,DuboisViolette,HKS1}.
Their results can be stated as follows.
Let the variables $\chi_0$ denote collectively
(i) the Yang-Mills field
strengths, their covariant derivatives
$D_{\alpha_1}\ldots D_{\alpha_k} F^a_{\mu\nu}$, the
antifields
and their co\-va\-riant
de\-ri\-va\-tives $D_{\alpha_1}\ldots  D_{\alpha_k}
A^{*\mu}_a$,$D_{\alpha_1}\ldots D_{\alpha_k}
C^{*}_a$; these transform according to 
the adjoint representation; and
(ii) the free 2-form field strengths
$H^0_{\mu\nu\rho}=(dB)_{\mu\nu\rho}$, their derivatives,
the antifields $B^{* \mu \nu}$,  $\eta^{* \mu}$,
$\rho^*$, their derivatives and the undifferentiated
ghost of ghost $\rho$. 
Then the representatives of $H(\gamma)$ can
be written as
$a=\sum_J \alpha_J(\chi_0){\omega}^J(C^a)$, where
the $\alpha_J(\chi_0)$ are invariant polynomials in
the $\chi_0$ 
and where the ${\omega}^J(C^a)$
constitute a basis of the Lie algebra
cohomology of the Lie algebra of the gauge group.  The
$\omega^J$ are
polynomials in the so-called ``primitive forms", i.e $tr
C^3, tr C^5 \hbox{ if } tr C^5 \not = 0$, etc. [For
instance, for $SU(3)$, the $\omega^J (C^a)$ can be taken
to be $\{1,$ $tr C^3,$ $tr C^5,$ $tr C^3$ $tr C^5\}$].

When the Chern-Simons coupling is turned on $(\lambda\not
= 0)$, the results are very similar but there are however
two modifications: (i) one must replace in the above 
cocycles
the free
field strengths $H^0_{\mu\nu\rho}$ and their derivatives by
the modified invariant field strengths $H_{\mu\nu\rho}$
(\ref{Hmunurho}) and their derivatives (we shall denote 
the new set 
of improved variables defined in this manner by $\chi$); 
(ii) the
ghost of ghost $\rho$ and the primitive form
$tr C^3$ drop from the cohomology since these elements now 
obey
the relation
$\gamma
\rho=\frac{\lambda}{3} tr C^3$, which indicates that $tr
C^3$ is exact, while $\rho$ is no longer closed. This last
feature underlies the Green-Schwarz anomaly cancellation
mechanism. 
We thus have:
\begin{theorem}
The representatives of $H(\gamma)$ can be written,
\begin{equation}
a=\sum_J \alpha_J(\chi)\overline{\omega}^J(C^a),
\end{equation}
where the $\chi$ stand for the field strengths 
($F^{a}_{\mu \nu}$
$H_{\mu \nu \rho}$),
the antifields ($A^{* \mu}_a$, $C^*_a$, $B^{* \mu \nu}$,
$\eta^{* \mu}$, $\rho^*$) and their covariant
derivatives, and $\overline{\omega}^J(C^a)$ is the subset
of the $\omega^J (C^a)$ which does not depend on $tr C^3$.
\newline \indent{\rm [So, for $SU(3)$, the
$\overline{\omega}^J(C^a)$ are just 
$\{1,tr C^5\}$]}.
\label{Hgamma}
\end{theorem}

\section{Antifield Dependence}
\setcounter{equation}{0}
\setcounter{theorem}{0}
The local cohomology $H^n_g(s \vert d)$ in  maximal form
degree $n$  - the only one considered here -
and ghost number $g$ is obtained by solving the
Wess-Zumino consistency condition,
\begin{equation}
s a_g + db_{g+1}=0,
\label{WessZum}
\end{equation}
where $a_g$ and $b_{g+1}$ are respectively local $n$- and
$(n-1)$-forms.
One must further identify solutions which differ by
$s$-exact and
$d$-exact terms (trivial terms), i.e,
$a_g \sim a'_g = a_g +s n_{g-1}+dm_g.$

Possible anomalies are elements of $H(s\vert d)$ for
$gh=1$ and counterterms correspond to $gh=0$. Furthermore,
if a counterterm is independent of the antifields, then
the gauge symmetries are preserved when this counterterm
(``deformation") is added to the action
\cite{BarnichHenn,GomisWeinberg}.

Our strategy for investigating the
Wess-Zumino consistency condition is identical to
the one used in \cite{BBH2} for the pure Yang-Mills case.
Any solution
$a$ can be decomposed according to the antighost number,
$a=a_0+\ldots+a_k$,
$antigh(a_k)=k$. 
In antighost
$k+1$, Equation (\ref{WessZum}) reads: $\gamma a_{k} +d
b_{k}=0$. Just as in \cite{BBH2}, it is easy to see that by
an allowed redefinition of $a$ one can choose $a_k$ such
that $\gamma a_k=0$ and $b_k=0$ if $k>0$. By theorem {\bf
\ref{Hgamma}} we thus have
$a_k=\sum_J \alpha_J(\chi)\overline{\omega}^J(C^a).$ Next
one shows that $\alpha_J(\chi)$ has to obey
the equation $\delta \alpha_J + d\beta_J=0$ and so defines 
a cycle
of the invariant characteristic cohomology. Were
$\alpha_J$  a trivial solution, i.e, $\alpha_J=\delta
\mu_J + d\nu_J$, then $a_k$ could be eliminated by an
allowed redefinition of
$a$. The obstructions to the removal of the antifields are 
therefore
elements of the (invariant) characteristic cohomology, 
which describes the
conservation laws of the theory \cite{BBH1}.

\subsection{Invariant characteristic cohomology}
The ordinary characteristic cohomology $H(\delta \vert
d)$ and the invariant characteristic cohomology have
already been separately studied in detail in antighost 
number
$>1$ for the Yang-Mills case and
for the p-form case
\cite{BBH1,BBH2,HKS1}. For the Yang-Mills case, both
cohomologies have been shown to vanish;
for a 2-form, they are given by $\rho^*$, which has
antighost $3$. 
The corresponding conservation law reads
$\partial_\mu \partial^{[\mu} B^{\nu \rho]} \approx 0$.
For the coupled system, one can verify that the
ordinary cohomology for $k>1$ is still given by $\rho^*$,
i.e., the coupling does not introduce any new cohomology
and does not remove $\rho^*$
\footnote{This follows from
the results of \cite{BBH1}, sections 10 and
11. 
Note that  the isomorphism
$H_k(\delta \vert d) \simeq H^{-k}(s \vert d)$ 
($k \geq 1$) determines
the local BRST cohomolgy at negative ghost number.
}. The result holds also for the invariant
cohomology:
\begin{theorem}
In antighost $k>1$, any invariant solution $\alpha_J(\chi)$
of the equation $\delta \alpha_J + d\beta_J=0$ can be
written,
\begin{equation}
\alpha_J=k_J\rho^* + \delta \mu_J + d\nu_J,
\end{equation}
where $\mu_J$ and $\nu_J$ are invariant and
where $k_J$ are constants.
\end{theorem}

\proof{The proof is based on a ``spectral sequence
argument" and works in the polynomial algebra
${\cal P}$ of spacetime forms with
polynomial coefficients. In the sub-algebra of gauge 
invariant
polynomial forms to which the invariant cochains belong,
$\delta$ can be split as
$\delta=\delta_{free}+\delta_{int}$ where $\delta_{free}$
increases by one the number of covariant
derivatives of the covariant
objects $\chi$ and corresponds to
$\lambda=0$, see (\ref{defdelta}). The invariant
characteristic cohomology
$H(\delta_{free}\vert d)$ for antighost number $>1$ is
given by $k\rho^*$.
The spectral sequence argument with filtration given
by the maximum number of covariant
derivatives shows then that
the invariant characteristic
cohomology is still given by
$\alpha_J=k_J\rho^* + \delta
\mu_J + d\nu_J$ when $\lambda$ does not vanish.  The
corresponding conservation law is of course
$\partial_\rho H^{\rho \mu \nu} \approx 0$.}

\subsection{Elimination of the antifield dependence at ghost
number zero and one}
We can now show that any solution of
(\ref{WessZum}) with ghost number equal to zero or one can 
be
assumed not to depend on the antifields.
Since $tr C^3$ is excluded from the
$\gamma$-cohomology, the ghost number of any
$\overline{\omega}_J(C^a)$ ($\not= 1$) is at least equal 
to $5$. In 
order to construct a solution of
(\ref{WessZum}) of ghost number $0$ or $1$ depending
non trivially on the antifields, we need an
element of the invariant characteristic cohomology which
is at least of antighost 5 or 4. But we have just shown
that the non trivial representatives of the invariant
characteristic cohomology with highest antighost number
are the multiples of
$\rho^*$, whose antighost number is 3. Therefore one cannot
find an $a_k$ with
$k\geq 1$ which yields a solution of
(\ref{WessZum}) depending truly on the antifields. We have 
thus proved:
\begin{theorem}
In each class of $H(s\vert d)$ of ghost 0 or 1,
there is an antifield independent representative ($\lambda 
\not= 0$).
\label{invd}
\end{theorem}

It is instructive to illustrate this theorem in the case 
of the
deformations of the solution of the master equation 
obtained by
varying the coupling constant  $\lambda \rightarrow
\lambda + \delta \lambda$.  Since such deformations
are consistent, they define elements of $H^0(s\vert d)$
\cite{BarnichHenn}.  In the present case, the deformation
reads explicitly
\begin{equation}
\delta \lambda [-\frac{1}{6} H^{\mu \nu \rho} 
\omega^3_{\mu \nu \rho}
+B^{*\mu\nu} \omega^2_{\mu \nu} + \eta^{*\mu} \omega^1_\mu +
\frac{1}{3} \rho^* trC^3].
\end{equation}
For  $\lambda = 0$, the antifield dependence is 
unremovable:  the
last term cannot be eliminated since $trC^3$ is non trivial.
However, if $\lambda \not= 0$, the antifield dependence
should be removable according to the theorem.  
And indeed, one easily verifies that the above BRST 
cocycle (modulo $d$)
is in the same class as
\begin{equation}
-\frac{1}{6} \frac{\delta \lambda}{\lambda} H^{\mu \nu \rho}
H_{\mu \nu \rho} \label{HH}
\end{equation}
since $\lambda \omega^3_{\mu \nu \rho} = H_{\mu \nu \rho} 
- \partial_{[\mu}
B_{\nu \rho]}$ and $H^{\mu \nu \rho} \partial_{[\mu}B_{\nu 
\rho]}
\approx \partial_\mu (H^{\mu \nu \rho} B_{\nu \rho})$.  
The representative
(\ref{HH}) does not involve the antifields.

\subsection{Antifield-dependent cohomology}
The analysis of the previous section relies crucially upon 
the
assumption that the ghost number is equal to $0$ or
$1$.  There exist 
cocycles (in form degree $n$) involving non trivially the 
antifields
when the ghost number is $\geq 2$. 
These cocycles  fall into two classes.  
(i) Solutions of type I involve the cocycle $\rho^*$ as 
term of
highest antighost number and are associated
to the third order conservation law $\partial_\rho H^{\rho 
\mu \nu}
\approx 0$.   (ii) Solutions of type II involve 
cocycles of antighost number $-1$
as terms of
highest antighost number and are associated 
to rigid symmetries (ordinary conservation laws).
\begin{itemize}
\item{Solution of type I}
\begin{eqnarray}
a&=&a_0 +a_1+a_2+a_3 \\
 &=&k_J ( \tilde{H} \overline{\omega}^{'''J}(C^a) +
\tilde{B}^*
\overline{\omega}^{''J}(C^a) + \tilde{\eta}^*
\overline{\omega}^{'J}(C^a)+
\tilde{\rho}^* \overline{\omega}^{J}(C^a)).
\end{eqnarray}

\item{Solutions of type II}

Let $a_\Delta = X_{\mu\nu\Delta} B^{*\mu\nu}+
Y_{\mu\Delta}A^{*\mu}$ be a complete set of invariant
representatives of $H^n_1(\delta \vert d)$. The $a_\Delta$
can be identified with the non-trivial global symmetries
\cite{BBH1} of the action (\ref{Lagrangian});  they
satisfy,
$\delta a_\Delta + \partial_\mu j_\Delta^\mu =0,$
where the $j_\Delta^\mu$ form a complete set of non-trivial
conserved currents \cite{BBH1}. The solutions of type II
can then be written,
\begin{eqnarray}
a&=&a_0 +a_1\\ \nonumber
&=&k_J^\Delta (
(-)^{\epsilon_{\overline{\omega}^{J}}+1}j^\mu_\Delta
\overline{\omega}_\mu^{'J}(C^a)
\\ &&+(X_{\mu\nu\Delta}
B^{*\mu\nu}+Y_{\mu\Delta}A^{*\mu})\overline{\omega}^{J}(C^a)
).
\end{eqnarray}
\end{itemize}

In the above formulas, the \ $\tilde{}$ \ denotes the 
form-dual
including appropriate multiplicative factors such that,
\begin{eqnarray}
\delta \tilde{B}^* + d\tilde{H}=0, \;
\delta \tilde{\eta}^* + d\tilde{B}^*=0, \;
\delta \tilde{\rho}^* + d\tilde{\eta}^*=0.
\end{eqnarray}
The $\overline{\omega}^{'J}
,\overline{\omega}^{''J},\overline{\omega}^{'''J}$
are obtained by lifting the $\overline{\omega}^{J}$ from 
the bottom of the descent equation,
\begin{eqnarray}
\gamma \overline{\omega}^{'''J} + d
\overline{\omega}^{''J}=0, \;
\gamma \overline{\omega}^{''J} + d
\overline{\omega}^{'J}=0, \;
\gamma \overline{\omega}^{'J} + d
\overline{\omega}^{J}=0, \;
\gamma \overline{\omega}^{J} =0.
\end{eqnarray}
The descent exists for all the $\overline{\omega}^{J}$
because these forms do not depend on $tr C^3$
\cite{DuboisViolette1}.

\section{Antifield independent solutions}

\subsection{Invariant cohomology of $d$}
In order to examine the antifield independent solutions of
(\ref{WessZum}) we need the following result on the
invariant cohomology of $d$:
\begin{theorem}
Let $a$ be a polynomial in the field strengths and their
(covariant) derivatives. Assume  the form degree of $a$
to be strictly smaller than $n$ and  $a$ to be $d$-closed:
$da=0$. Then one has $a=\overline{P}(F^a) + db$, where $b$
depends only on the field strengths and their
(covariant) derivatives and
$\overline{P}$ is an invariant polynomial in the forms
$F^a$ which does not contain the quadratic invariant 
$tr F^2$.
\end{theorem}
\proof{When $\lambda=0$, the invariant cohomology of $d$ is
given by the invariant polynomials in the 3-form $H$ and the
2-form $F^a$. That is, any solution of $da=0$ which depends
only on the field strengths and their derivatives can be
written
$a=P(H,F^a)+ db(H_{\mu\nu\rho},\partial_\alpha
H_{\mu\nu\rho},\ldots, F^a_{\mu\nu}, \partial_\alpha
F^a_{\mu\nu},\ldots)$  (See
\cite{DuboisViolette} and \cite{HKS1}).

When $\lambda \not =0$, the invariant cohomology  of $d$
is given by the invariant polynomials in the $F^a$ which
do not depend on $tr F^2$. Indeed, $tr F^2$ becomes exact
in the algebra of {\em invariant} polynomials, $tr F^2
= (\lambda)^{-1} dH$.  Furthermore, 
$H$ disappears also from the cohomology since it is no
longer $d$-closed.}

\subsection{Results}
We can now work out the antifield independent solutions of 
the
Wess-Zumino consistency condition $\gamma a_g +
db_{g+1} = 0$.
These fall also into two classes.  The first one involves
the solutions for which the
$(n-1)$-form $b_{g+1}$ either vanishes or can be made to
vanish by redefinition. The second one involves
the solutions that lead to a non trivial
descent.  The solutions of the first class ($\gamma
a_g =0$) are easily
determined since we already know the cohomology of
$\gamma$.  We thus  focus  on the solutions of
the second class, associated to a non trivial descent,
\begin{eqnarray}
\gamma a_g + db_{g+1} = 0, \,
\gamma b_{g+1} + dc_{g+2} = 0, \,
\dots,
\nonumber \\
\gamma m_s + dn_{s+1} = 0, \,
\gamma n_{s+1} =0.
\end{eqnarray}
The last term $n_{s+1}$ in the descent is annihilated by 
$\gamma$
and thus takes the form $n_{s+1} = \sum_J P_J
\overline{\omega}^J(C^a)$ where $P_J$ is an invariant 
polynomial in
the field strength components $F^a_{\lambda \mu}$, 
$H_{\lambda \mu \nu}$
and their (covariant) derivatives.  The next to last 
equation
implies then $d P_J  =0$ and thus, by Theorem {\bf 5.3},
the polynomial $P_J $ is actually an invariant polynomial 
in the
forms $F^a$ which may be assumed not to involve $trF^2$
($H$ and $trF^2$ drop out).

This is exactly the form encountered in the pure 
Yang-Mills case
since the variables related to the 2-form $B_{\mu \nu}$ no
longer appear.  Accordingly, we may proceed along the 
lines of reference
\cite{DuboisViolette} to analyse
which cocycles
$\sum_J \overline{P}^J (F^a) \overline{\omega}^J(C^a)$
can be lifted all the way up to a solution $a_g$ of the
Wess-Zumino consistency condition in degree $n$.
We refer the reader to that work for the details.  The 
only difference
is that we start here with a  restricted form of the bottom
since it involves neither $tr F^2$ nor $tr C^3$.  Thus, 
all the solutions
of the second class containing $tr C^3$ or $tr F^2$,
or related to them through the descent, become
trivial with the introduction of the 2-form $B_{\mu \nu}$
($\lambda \not= 0$).

\subsection{Example in spacetime dimension $d=10$}
We shall illustrate the above procedure in the 
ten-dimensional case,
for ghost number zero and one.  For definiteness, we take 
the gauge
group to be $SU(n)$ ($n\geq 6$) so that the primitive
forms $trC^3$, $trC^5$, $trC^7$, $trC^9$ and $trC^{11}$ 
are all
independent.
As we have seen, the antifields drop out from the 
cohomology.

At ghost number one, the only solutions of the Wess-Zumino 
consistency
condition $\gamma a +db = 0$ are of the first type, 
$\gamma a = 0$
($b = 0$ by redefinitions).
This is because there is no non trivial bottom of the
descent with (ghost number $+$ form degree) equal to $10$.
Thus, the solutions are strictly invariant; they are  the
invariant polynomials in the individual components 
$F^a_{\lambda \mu}$,
$H_{\lambda \mu \nu}$ and their (covariant) derivatives.

By contrast, there are no strictly invariant solutions at 
ghost number
one and the only solutions of the Wess-Zumino consistency 
condition
are of the second type, associated to a non trivial descent.
The possible bottoms must have (ghost number $+$ form 
degree) equal
to $11$.  The only non trivial ones are $trC^{11}$ and 
$trF^3 trC^5$.
Both can be lifted all the way up to form degree $10$ and 
lead respectively
to the irreducible anomaly
\begin{equation}
a_{IRR} = Q^{10,1}
\label{exp1}
\end{equation}
and the factorizable one,
\begin{equation}
a_{F} = trF^3 Q^{4,1}
\label{exp2}
\end{equation}
where $Q^{10,1}$ is defined through $dQ^{10,1} + \gamma 
\omega^{11}_{CS} =0$,
$d\omega^{11}_{CS} = trF^6$ ($\omega^{11}_{CS}$ is the 
eleven-dimensional
Chern-Simons form) while $Q^{4,1}$ is the familiar 
Adler-Bardeen-Jackiw
anomaly in four dimensions ($dQ^{4,1} + \gamma 
\omega^{5}_{CS} =0$,
$d\omega^{5}_{CS} = trF^3$).  There is no factorizable 
anomaly related
to $tr F^2$ since these become trivial through the
coupling to the two-form (Green-Schwarz anomaly cancellation
mechanism).  Expressions (\ref{exp1}) and (\ref{exp2})
are the only solutions of the
Wess-Zumino consistency conditions at ghost number one for 
$SU(n)$
($n \geq 6$).  For other groups,  these solutions exist 
but may be
trivial if there is no irreducible three-index or 
six-index Casimir invariant.

\section{Conclusions}
In this letter, we have provided the general solution of the
Wess-Zumino consistency condition for the Chapline-Manton
model, for all ghost numbers and without use of power
counting (which would not help much in any case, since the
coupling constants are dimensionful).
The antifields have been explicitly included, but have been
shown not to bring in new solutions at ghost
numbers zero and one.

Our analysis has been carried out in the case of a simple 
Lie
group $G$ and for the quadratic Lagrangian (1.1).  
Since we have not used power counting, the results can be 
extended easily to 
higher-derivative gauge-invariant Lagrangians.  They can
also be extended 
to the case where $G$ is the direct product of simple 
groups by
$U(1)$ factors.   The simple factors can all
be treated as above (if one brings in a 2-form for
each such factor).  The analysis of the abelian factors is 
more complicated
since they can lead to antifield-dependent solutions
even at ghost number zero or one, but it proceeds exactly as
in \cite{BBH2}.


\begin{thebibliography}{100}
\bibitem{GS} M. B. Green and J. H. Schwarz, {\em Phys. 
Lett.}
{\bf 149B} (1984) 117.
\bibitem{GSW} M. B. Green, J.H.  Schwarz and E. Witten,
{\em Superstring Theory} (2 vols), Cambridge University
Press, Cambridge (1987).
\bibitem{ChaplineManton} G.F. Chapline and N.S. Manton,
{\em Phys. Lett.} {\bf B120} (1983) 105.
\bibitem{Nito} H. Nicolai and P.K. Townsend, {\em
Phys. Lett.} {\bf 98B} (1981) 257.
\bibitem{Cham} A. H. Chamseddine, {\em Nucl. Phys.}
{\bf B185} (1981) 403; {\em Phys. Rev.} {\bf D24} (1981) 3
065.
\bibitem{BergRooWitNieu} E. Bergshoeff, M. de Roo, B. de
Wit and P. van Nieuwenhuizen, {\em Nucl. Phys.} {\bf B195}
(1982) 97.
\bibitem{GomisWeinberg}
J.~Gomis and S.~Weinberg, {\em Nucl. Phys.\/} {\bf B469}
(1996) 473.
\bibitem{BBH2}  G. Barnich, F. Brandt and M. Henneaux,
{\em Commun.Math.Phys.} {\bf 174} (1995) 93.
\bibitem{BBH1} G. Barnich, F. Brandt and M. Henneaux,
{\em Commun.Math.Phys.} {\bf 174} (1995) 57.
\bibitem{BRST} C. Becchi, A. Rouet and R. Stora, {\em
Comm. Math. Phys.} {\bf 42} (1975) 127; {\em Ann. Phys.}
(NY) {\bf 98} (1976) 287.; I.V. Tyutin, {\em Gauge
invariance in field theory and statistical mechanics},
Lebedev preprint FIAN, n.39 (1975).
\bibitem{BV} I.A. Batalin and G.A.
Vilkovisky, {\em Phys. Lett.} {\bf B102} (1981) 27; {\em
Phys. Rev.} {\bf D28} (1983) 2567; {\em
Phys. Rev.} {\bf D30} (1984) 508.
\bibitem{HenneauxTeitelboim} M. Henneaux and C. Teitelboim,
{\em Quantization of Gauge Systems}, Princeton University
Press,
Princeton (1992).
\bibitem{FHST} J. Fisch, M. Henneaux, J. Stasheff and C.
Teitelboim,{\em Commun. Math. Phys.} {\bf 120} (1989) 379.
\bibitem{FH} J.M.L. Fisch and M. Henneaux, {\em Commun.
Math. Phys.}
{\bf 128} (1990) 627.
\bibitem{HenneauxCMP} M. Henneaux, {\em Commun. Math. Phys.}
{\bf 140} (1991) 1.
\bibitem{CohoH}  J.A. Dixon, {\em Cohomology and
Renormalization of Gauge Theories I, II, III}, Unpublished
preprints (1976-1979); {\em Commun. Math. Phys.} {\bf 139}
(1991) 495; G. Bandelloni, {\em J. Math. Phys}
{\bf 27} (1986) 2551; {\em J. Math. Phys} {\bf 28} (1987)
2775; F. Brandt, N. Dragon and M. Kreuzer,
{\em Phys. Lett.} {\bf B231} (1989) 223;
{\em Nucl. Phys.} {\bf B332} (1990) 250;
{\em Nucl. Phys.} {\bf B332} (1990) 224;
M. Henneaux, {\em Phys. Lett.} {\bf B313} (1993) 35.
\bibitem{DuboisViolette} M. Dubois-Violette, M. Henneaux, 
M. Talon
and C. M. Viallet, {\em Phys. Lett.} {\bf B267} (1991) 81.
\bibitem{HKS1} M.  Henneaux, B.  Knaepen and C.
Schomblond, hep-th/9606181, to appear in {\em Commun. Math.
Phys.}
\bibitem{BarnichHenn} G. Barnich and M. Henneaux,
{\em Phys.Lett.}
{\bf B311} (1993) 123.
\bibitem{DuboisViolette1} M.~Dubois-Violette, M.~Talon and
C.~M. Viallet, {\em Commun. Math. Phys.\/}
  {\bf 102} (1985) 105.
\bibitem{Poincare} P.J. Olver, {\em Applications of Lie
Groups to
Differential Equations}, Springer-Verlag, (1986).
\end{thebibliography}
\end{document}